\documentclass[12pt]{article}
\begin{document}

\Large
 
\center{Role of fluctuation, disorder and catalyst in
graphite-diamond transition}

\center{Akihito Kikuchi\footnote{Corresponding author.
\\Phone/Fax:(+81)471-36-3291~
e-mail:miare@issp.u-tokyo.ac.jp}
 and Shinji Tsuneyuki}

\center{Institute for Solid State Physics, University of Tokyo}

\center{Kashiwa-no-ha, 5-1-5, Chiba,277-8581,Japan}

\abstract{
The pressure-induced structural transition from graphite
to diamond is investigated by semi-empirical 
molecular dynamics simulation. 
The result shows that
the graphite-diamond transition is a cooperative 
process with large fluctuation. 
We  studied catalyst-aided effect 
by introducing a simple model 
into a conventional tight binding Hamiltonian of
carbon bonding structure. 
The obtained result suggest that 
weak disorder in graphite  
not only accelerate the transition to cubic diamond
but also will be an origin
of high-pressure polymorph of carbon, such as hexagonal diamond.
}

\large
\section{Introduction}
Carbon has various meta-stable polymorph, such as diamond,
graphite, fullerene, or nanotube and so on.
The structural transition between these polymorph structures,
often induced by high pressure, has been intensively 
studied for years. 
For example, 
diamond is obtained from graphite by static or shock-wave
compression experiments\cite{BUNDY_KASPER,CLARKE}.
The diamond obtained in such ways shows several distinct phases,
such as, cubic diamond(CD) and hexagonal diamond(HD),
which have different stacking and
networking structures of sp$^3$ tetrahedrons.
In many cases, compressed
graphite directly turns into cubic diamond with the aid of catalyst,
accompanied with the change in stacking
from so-called AB type  into ABC one\cite{1STP_CAL}.
However, HD phase has also been 
synthesized by static compression from
well-crystallized graphite\cite{BUNDY_KASPER} or
shock-compression\cite{CLARKE},
in which AB staking is kept after quenching.
However, in the compression of polycrystalline graphite,
no hexagonal phase is obtained\cite{POLYGRA}.
It is also reported that the transformation to HD occurs
under quasi-hydrostatic pressure, but only CD phase is quenched\cite{YAGI}.
Under certain conditions, 
"n-diamond" phase also forms under compression \cite{POLYGRA,N_DIA}.
There are several theoretical studies for graphite-diamond
transition from first principles.
Scandolo et al. had simulated the transformation
from graphite to diamond\cite{1STP_CAL}. 
Their simulation has created both CD and HD structure.
Tateyama et al. investigate the activation
barriers and the transition paths in the transition to cubic or
hexagonal diamond from graphite 
and discussed on the origin of diamond polymorphs,
which is attributed to the sliding movement of graphite layers
under compression\cite{G2D}.
However, from the computational limit, the
ab-initio simulations use at most tens of atoms 
and they may not be sufficient for describing
some aspect of diamond transition, since
such a crystal growth process may be subject to
large scale structural fluctuation.
The small size simulation is likely to ignore that 
the graphite-diamond transition
shall essentially be a cooperative phenomena 
caused by  local re-combination of bonding structure.
In addition, the realistic mechanism
in choosing HD or CD structure under compression
has not been clarified enough by those ab-initio studies.
In the present work we adopt a tight-binding simulation
using hundreds of atoms in order to study
possible effects in the transition, such as
fluctuation, disorder, or catalyst.
At first we simulate the cooperative
process of the graphite-cubic-diamond transition.
Next, the simulation for  catalyst
effect in the transition to high-pressure
polymorphs of diamond is given.
There a model for catalyst-carbon combination is included into the
TB Hamiltonian for carbon bonding structure.
We propose a catalyst-aided-mechanism
in the transition to carbon polymorphs under pressure.

\section{Methodology}

In the simulation of pressure-induced 
structural transition, we adopt a
constant-pressure scheme proposed by
Wentzcovitch et al.\cite{Wentz}.
In the evaluation of total energy and force in molecular dynamics,
we use tight binding Hamiltonian 
for hydro-carbon system proposed by Winn et al.\cite{PARAM_CH}. 
Throughout the present work, the time interval of
one MD step is set to be 1fs.

In the simulation, we start with graphite structure at 0Pa.
The initial stacking of graphite is set to be AB type.
In case of the simulation for catalyst-aided-effect,
the catalysts are randomly distributed
between graphite sheets as intercalated atoms.
The pressure is raised to the desired value
at the rate of 1GPa/1MD steps and kept constant after that.
The temperature is also kept constant.
The simulation in this way does not cause 
any kind of abrupt change in the cell dynamics.
In the constant pressure simulation\cite{Wentz}, 
we set the virtual mass assigned to cell deformation
heavy enough so that the 
inner stress shall go parallel with the 
increase in the external pressure.

\section{Simulation of graphite-diamond transition}
In this section, the result of graphite-diamond transition
is shown.  

In the present work, the graphite
directly turns into cubic diamond at 150-200GPa and 5000-6000K.
However, hexagonal phase is not obtained.

Figure 1 shows  the growth of interlayer bonding
between two neighboring graphite sheets.
The compression process is executed at 200GPa, 6000K.
The unit cell contains 12 graphite sheets with AB stacking and
each graphite sheet has 40 carbon atoms.
The development of interlayer bonding is estimated by the
ratio defined as follows.

\begin{equation}
R_{i,i+1}= \frac{2\times
\mbox{(Number of atoms with interlayer bonding between i and (i+1)-th layer
)}}{\mbox{ Number of atoms in a layer}}
\end{equation}

If this ratio is 0, there is no bonding structure between the
corresponding two layers.
On the other hand, if this ratio becomes 1,
the two sheets are perfectly glued with each other
by interlayer bonding.

The transition process is described as follows.
When the structural fluctuation has grown enough,
the diamond structure grows  very rapidly.
The development of diamond structure
along the c-axis is slower than that goes parallel
with the original graphite plane.
The domain boundaries of the diamond structure
come to play a role of the mold so that  they
combine with nearby graphite sheet.
In this way, the diamond structure will grow as a collective phenomenon.
Even if the more wide graphite sheets are used
in the unit cell of the MD simulation, the transition occurs similarly
in the same range of the temperature and the pressure.

\section{Catalyst-aided transition}
In this section, the role of catalyst is investigated
by the simulation including catalyst atoms,
which are randomly distributed between graphite planes
as intercalated atoms.
In order to simulate the  catalyst effect,
we make use of a model,
where the only s-orbital of catalyst interacts with carbon.
As the tight binding parameters between carbon and catalyst, 
we start from those between carbon and hydrogen. 
We introduce a parameter $\lambda$ and construct
the model of  catalyst-carbon coupling
by multiplying $\lambda$ to matrix elements and repulsion terms 
between hydrogen and carbon\cite{PARAM_CH} in the following way.

$t_{sp\sigma} \Rightarrow \lambda t_{sp\sigma}$ 

$t_{sp\pi} \Rightarrow \lambda t_{sp\pi}$ 

$t_{ss\sigma} \Rightarrow \lambda t_{ss\sigma}$ 

$V_{\mbox{rep}} \Rightarrow \lambda V_{\mbox{rep}}$ .

The site energy for s  orbital is fixed to that of hydrogen.

The dependence of the cohesive energy and the bond length
on $\lambda$ in a C$_2$H$_6$ like molecule is given in figure 2.

The transition process from graphite to diamond
proceeds at lower pressure and temperature with the aid of catalyst.
In the present simulation, 
the transition temperature to diamond is 150-200GPa
in the compression of genuine diamond.
However, if the catalyst is induced,
at most 4-6 catalyst atoms in 240 carbon atoms,
and even if the catalyst-carbon coupling is not so strong ($\lambda \sim 0.6$),
the  transition pressure is decreased below 100GPa. 
It will be attributed to
the catalyst-induced buckling of
the graphite basal plane,
in which $\pi$ orbitals are activated[Figure 3].

\section{Formation mechanism of hexagonal diamond}
A kind of stable diamond structure called as hexagonal phase
is sometimes obtained by shock compression and static compression.
This phase has the structural property
very similar to normal cubic diamond phase.
The first principle calculation shows that 
the cohesive energy in hexagonal phase is slightly higher
than that of  cubic phase\cite{G2D}. 
In the TB model in the present work,
the difference in cohesive energy between CD and HD
is also very small. From the energetic viewpoint,
it is also possible for hexagonal phase to be formed
in the compression of graphite. However, the simulation
in the former sections, generated only cubic phase.
It will be attributed to the difference
of the height of activation barrier[Figure 4]
between  the transformation paths from graphite to 
cubic phase and hexagonal phase,
as is demonstrated by a first principles calculation \cite{G2D}.

According to the molecular dynamics simulation in the present work,
under the presence of catalysts, the hexagonal phase is often formed,
when we set certain types of initial geometry.
The figure 5 shows one of the results(150GPa,6000K,$\lambda=1.2$).
The transition path to diamond phase is postulated in the following way.
When cubic diamond is formed, the stacking 
should change from AB stacking of graphite
into ABC one, as graphite sheets slide collectively.
On the other hand, in hexagonal phase, the AB stacking must be kept.
In the  geometries under the presence of impurities,
the collective slide may be  disturbed and "pinned" 
by the coupling between catalyst and carbon[Figure 6].
Then the short range order of hexagonal 
diamond is formed around the catalyst 
and it finally covers up whole crystal.

Figure 7 and 8 show the simulated result for a
certain initial geometry with impurities.
Figure 7 shows the synthesized phases after the compression
as long as 2ps (2000MD steps)
with various conditions for $\lambda$, 
and the temperature and the pressure.
The simulations are started from the same initial geometry.
In figure 8,
in order to investigate microscopic mechanism
of the transition, the process of interlayer bonding
with different conditions is shown, using
plots similar to figure 1. 
Fig.8(a)-(c) show the formation of synthesized phases in Fig.7.
The common initial configuration of impurities
used in  the simulation in Fig.7 and Fig.8 is given in Fig.8(d).
In the simulated result, it seems that there is a certain tendency
with respect to $\lambda$, temperature and pressure in the transition. 
In a certain range of $\lambda$ and the temperature,
the transition to hexagonal phase occurs.
However, if the temperature is set to be higher 
while  $\lambda$ is fixed,
the hexagonal phase tends to vanish.
Instead, the transition to cubic phase occurs.
When the value of $\lambda$ becomes large,
the graphite turns into cubic phase,
but the long range order is often damaged
and large defects remain.
The result of Fig.7 and Fig.8  is obtained by a set of test-run
and it does not necessarily  mean the phase diagram itself.
However,  if  the different condition, such as
the different initial velocity of each atom, 
is given, the simulation gives similar tendency
when the similar initial configuration is used.

From Fig.8, the scenario of graphite-diamond transition 
induced by the competition of several physical conditions,
i.e., temperature, pressure, presence of impurity, and so on,
is described as follows.

\begin{itemize}
\item In the case of genuine graphite

In this case, the formation of interlayer bonding proceeds
rapidly in the direction parallel with graphite plane,
while it goes slowly along c-axis.
In this case, the transition path to cubic phase 
having the lower activation barrier will be selected[Figure 4].
The simulated result shows the it costs
long time before the formation of diamond structure
begins. It will be interpreted to be the
time lag until the compressed system
will find the saddle point in the potential surface.

\item Under the presence of catalyst

Since the presence of catalyst causes the
combination with carbon and the buckling of graphite plane,
the formation of interlayer  bonding begins around impurities
in earlier stage of the reaction[Figure 8(a)].
However, the catalyst is a kind of disorder and it
somewhat disturbs the formation step of interlayer bonding
and causes the fluctuation in the cooperative transition 
toward cubic phase, letting down the transition speed[Figure 8(b)].
Owing to staggering in the layer-by-layer transition steps,
there arises  possibility
of the  transition to different
carbon polymorph other than cubic diamond.
In addition, if the sliding movement may be hindered and "pinned"
by the weak coupling between catalyst and carbon,
the graphite AB stacking is locally preserved, 
from which local core of hexagonal structure
forms and finally covers the whole crystal[Figure 8(b)].
In fact, the transition to HD induced by impurities
are severely dependent on randomness
with respect to initial configurations of impurities and
it does not always occur in the simulations like this.

\item Catalyst in higher temperature

In case of too high temperature,
as the structural fluctuation of graphite sheets will exceed 
the catalyst-induced pinning in graphite sliding,
the local core of hexagonal diamond may not be formed[Figure 8(c)].
Thus cubic diamond will form, accompanied with strain change.

\item With large catalyst-carbon interaction $\lambda$

When the $\lambda$ is set to be larger,
the graphite sheets are glued firmly  by impurities.
Then the fluctuation in the cooperative step
in the transition does not become large
and the compressed graphite is rapidly 
stabilized into cubic phase.
\end{itemize}

In fact, in the actual conversion process,
the crystal will include various structural defects 
other than intercalated impurities.
In the neighborhood of defects, atoms can move more freely
and it will be possible for local hexagonal
diamond structure to be formed out of structural fluctuation.
In the simulation where the points defects are included,
the short-range order similar to hexagonal phase is actually obtained.

\section{Comparison to pressure-induced transition of fcc fullerene}
It is demonstrated that the fullerene fcc crystal 
turns into  transparent glassy chips 
of amorphous diamond phase by shock compression\cite{HIRAI}.
This phase is constructed from
short range order of sp$^3$ bonding
and it is distinguished from normal
amorphous phase of carbon. In general,
amorphous carbon is interpreted to be disordered phase of graphite.
By MD simulation, the pressure-induced amorphization
in fullerene can also be pursued. 
According to  the MD result, 
fcc fullerene collapses easily above 50GPa, 2000K 
and turns into amorphous phase.

Figure 9 shows the electronic structure 
of the finally fabricated material after the compression.
These figures show the case of the following conditions, 
(a)65GPa,3000K,(b)125GPa,5000K,
(c)125GPa,5000K with presence of 30Hydrogens.
The simulation uses
 240 carbon atoms , i.e., 4 C$_{60}$ molecules in a unit cell.
In these figures, real and dotted lines
show the total DOS and the contribution from threefold carbon, 
respectively.
The rough feature of DOS is different to
that of initial fcc fullerene\cite{C60FCCBAND} and it is
similar to that of diamond, except that
gap states contributed from dangling bonds  remain
between valence and conduction band.
If the contribution from dangling bonds 
between valence and conduction bands
is large, the fabricated material will
lose the transparent optical property 
and it will not be qualified
to be called amorphous diamond.
However, a comparison between figure 9(a) and 9(b) suggests
that the contribution from dangling bonds
will be reduced after the compression
at sufficiently higher temperature and pressure.
(In case of figure 9(a), the ratio 
of fourfold atom is at most 50 \%.
In case of figure 9(b), the ratio 
of fourfold atom amounts to 80 \%.)
The shaded zone in Fig.9(b) corresponds
to the conduction-valence gap of normal diamond structure.
However, gap states are still left in it.
In fact, there may be other mechanism
that will reduce the contribution from dangling bonds.
For example, it is well-known that the presence of hydrogen atoms
reduces the number of dangling bonds in case of amorphous silicon.
Figure 9(c) shows the DOS under the presence of hydrogen.
In this case the ratio of fourfold carbon amounts to 60\%.
According to the simulations in the similar way,
the number of dangling bonds decreases as
the more hydrogens are doped. 
However, even if hydrogens are included,
clear valence-conduction gap does not
open because of dangling bond states.
It is unlikely that a small quantity of impurities
will behave as catalyst and enhance 
the transition to amorphous diamond,
as we have seen in the previous section
for the case of graphite-diamond transition.
In the transition of bonding state from sp$^2$ to sp$^3$,
it seems that higher pressure
plays a more relevant role than the presence of 
hydrogen does. If the initial
configuration with hydrogen, which is same as in Fig.9(c),
is compressed by higher pressure (above 100GPa),
the number of fourfold atom does not increase so much
compared to Fig.9(a) of carbon only case.

In case of the compression of C$_{60}$,
the collapsed fullerene turns into amorphous phase.
Since there are number of quasi-stable configurations in amorphous phase,
the transition toward diamond-like phase is very slow
and the system is apt to be trapped
such  transient phases by quenching.
As an example,  figure 10 shows 
the two reaction path noted as (A) and (B).
In both path, the pressure is increased 0Pa to 125GPa
in the interval from 0.0ps to 0.5ps in the figure,
and compression is executed adiabatically in the first stage. 
The temperature increases from initial 300K to
about 2500K at the turning point of two paths.
In path (A), the simulation is executed furthermore adiabatically
and the temperature finally increased to 5000K.
In path(B), in which the temperature is kept to 2500K,
the transition to sp$^3$ bonding becomes slow due
to lower temperature.  Taking after different paths of(A) and (B),
the system is trapped by different semi-stable configurations.

The presence of too many intermediate phase 
between fullerene and  diamond  will be a major difference
to pressure-induced graphite-diamond transition.
Graphite and diamond  are located in very close
configurations in the potential surface and there is 
no stable transient state in the transition path from 
genuine graphite to genuine diamond.
Therefore graphite rapidly transforms into the cubic diamond
without being trapped by any quasi-stable structure.
Thus perfect diamond phase remains alone after the quenching
and, in general, amorphous diamond phase will not be
obtained in graphite-diamond transition.
Concerning this, the role of catalyst 
in graphite diamond transition can also be stated in this way.
The presence of impurities induces
certain quasi-stable structures in the transition path from graphite. 
In the present simulation,
such a quasi-stable structure corresponds
to the local core of hexagonal diamond.
Since such a quasi-stable structure 
staggers the transition steps from graphite,
there arises possibility of transition
to carbon polymorph other than cubic diamond.

\section{Discussion}
The present work is likely to account for some aspect 
of non-equilibrium process in graphite-diamond transition realized 
by shock compression. In shock compression, the high pressure phase,
which may not have arrived equilibrium state, 
is quenched and frozen, since  the temperature decreases very rapidly
while the high pressure is still kept.
In the present work, the hexagonal phase is generated
as a result of non-equilibrium core formation around impurities,
and such a HD phase is also stable and can be quenched.
However, if the high temperature environment continues
sufficiently long
as is realized in the actual conversion process,
and if the system will finally reach the equilibrium,
the system will select cubic phase as a equilibrium phase,
since cubic phase is slightly stabler than hexagonal one\cite{G2D}.

Several reports has demonstrated  that the
hexagonal diamond phase often grows in the compression 
of well-crystallized crystal\cite{CLARKE,YAGI}. 
It is interpreted that the hexagonal phase
is formed by the inhibited collective slide of graphite,
which is caused by the difficulty of sliding movement
in well-crystallized graphite\cite{G2D}.
However, the present work
suggests that the well-crystallized property itself is not
the direct origin of hexagonal phase.
It would rather be attributed to weak randomness in the crystal.
The presence of disorders disturbs the collective
step of the transition and gives birth to 
the growth of the short-range order of hexagonal phase.
The well-crystallized structure of  graphite helps 
such a short-range phase easily turn into a long-range order.
  
The report of the growth of hexagonal phase by static compression  
at low  temperature\cite{YAGI}
will be interpreted as the catalyst-induced
"pinning effect" to collective slide in graphite sheets. 
In that case, the formation of local core of hexagonal phase
is not disturbed by the thermal fluctuation due to low temperature.
Therefore the collective slide
is inhibited and the hexagonal phase is allowed to extend.
However, in the present work,
the hexagonal phase does not generate in 
the picosecond-order simulation at low temperature.
Concerning with graphite-diamond transition,
Yagi et al. also reported  that,
when the graphite is heated by YAG laser under pressure,
cubic diamond is formed instead of hexagonal phase\cite{Utsumi}.
This phenomena will be explained by the present simulation.
Because of too high temperature, impurities does not 
suppress the sliding movement of graphite and
the local core of hexagonal phase will not be formed.

The transition temperature and pressure
in the present work are higher than 
those in the actual experiment of graphite-diamond transition
($\sim$20Pa, $>$2000K).
One of the reason is as follows.
The present simulation uses the perfect crystal of graphite.
But the actual graphite crystal has
various structural defects from which
interlayer bonding will easily grow.
In addition, the present simulation
does not follow the shortest path from graphite to diamond.
The compressed graphite 
drifts and fluctuates on the potential surface,
and it can transform the structure
only if the saddle point is find.
This is different to the simulation
chasing after the shortest transition path
as was given in Ref.\cite{G2D}.
In the present work,  large fluctuation should be grown for 
the compressed graphite to cross the saddle point
in a finite, possibly very short, simulation time.
If we use much longer simulation time,
it is likely that the transition occurs
in lower pressure and temperature.
It may be possible that the 
tight-binding parameters are not accurate, since they
are obtained by the fitting to 
graphite and diamond structure, but not fitted
to the intermediate phase through transition.
However, the energy difference between the
intermediate phase and the cubic diamond
is evaluated as about 0.5-0.6 eV, which is 
no so different to the first 
principle result ($\sim$0.4eV)\cite{G2D}

An ab-initio calculation in the reference of \cite{LI_INT}
suggests  that, in the compression of
heavily doped GIC( LiC$_{12}$),
hexagonal diamond forms in a way such that carbon cages
will hold lithium in their centers.
In it the c-axis of the generated hexagonal diamond 
is parallel to that of original GIC.
The periodic distortion in graphite planes
by intercalated atoms in heavily doped LiC$_{12}$ appears to be
the key to diamond transition under pressure.
However, such a  transition path to
hexagonal phase is essentially different to that
suggested by the present work.
In the present work, the c-axes of synthesized
hexagonal diamond and those of the initial graphite
are perpendicular, as are demonstrated 
in several experiments of hexagonal phase formation\cite{CLARKE,YAGI}.
In addition, the present simulation treated the case where 
the doping density of intercalated atom is
much smaller than that of Ref.\cite{LI_INT}.
The present work suggests that randomness with respect to the
initial configuration of impurities plays an essential role.

\section{Summary}
In the present work, we simulated
diamond transformation from graphite.  
By compression, local core structures 
of diamond are formed in graphite at first.
The transition proceeds collectively, accompanied with fluctuation,
as the local diamond domain grows its volume
by turning graphite sheets into the new boundary.
In the present simulation,
the genuine graphite turns only into cubic diamond,
but not into hexagonal one. 
This will be attributed to the presence of higher 
activation barrier from graphite to hexagonal phase.
The pressure-induced structural transition
from graphite is subject to
the presence of disorder, such as catalyst
and possibly structural defect. 
We used a model which mimics the catalyst-aided diamond
transformation from graphite.  
One of the role of the catalyst is
the large reduction of transition pressure.
Since the combination of carbon and catalyst causes
the buckling of graphite plane where
$\pi$ electrons are activated, 
the interlayer bonding is easily formed there.
At the same time, catalyst disturbs the steps of
the rapid collective transition procedure into cubic diamond.
In some cases, owing to impurities,
the collective slide of graphite layers
is prohibited and hexagonal diamond structure is locally formed,
from which the whole cell turns into hexagonal diamond.
The simulated result suggest that 
"weak disorder" will be one of the origin of well-crystallized
carbon polymorph such as hexagonal phase. 

\section*{Acknowledgment}
The computation in this work has been done
using facilities of the supercomputer center,
Institute for solid state physics, University of Tokyo.

\section*{Figure captions}

\begin{itemize}

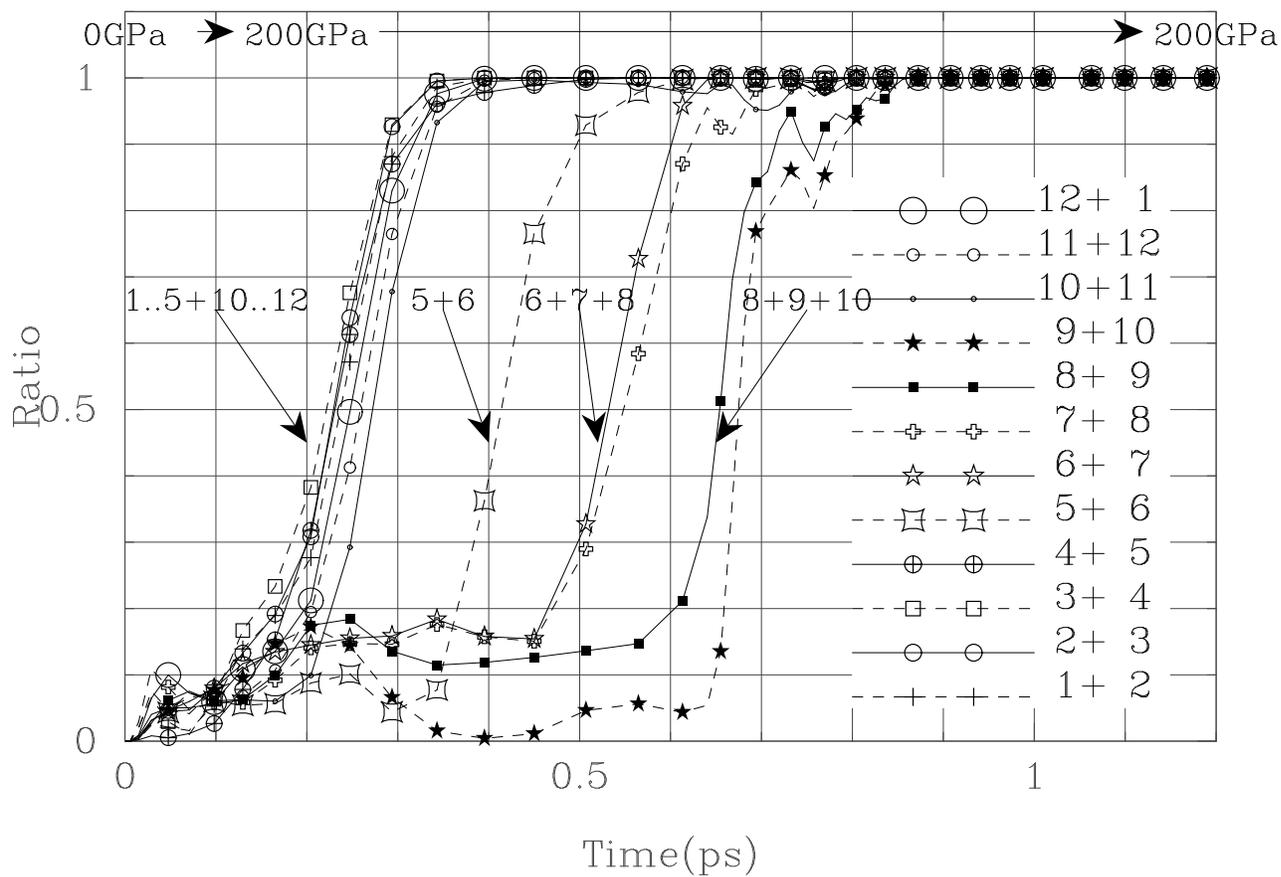
\item Figure 1:  Interlayer bonding formation in GR-CD  transition.
                 The graphite sheets are indexed as 1,2,..12
                 from bottom to top in the unit cell with
                 periodic condition.
                 The notation, for example, "1+2",
                 means the interlayer bonding formation
                 between 1st and 2nd graphite layers and so on.
                 One graphite sheet contains 40 carbon atoms.

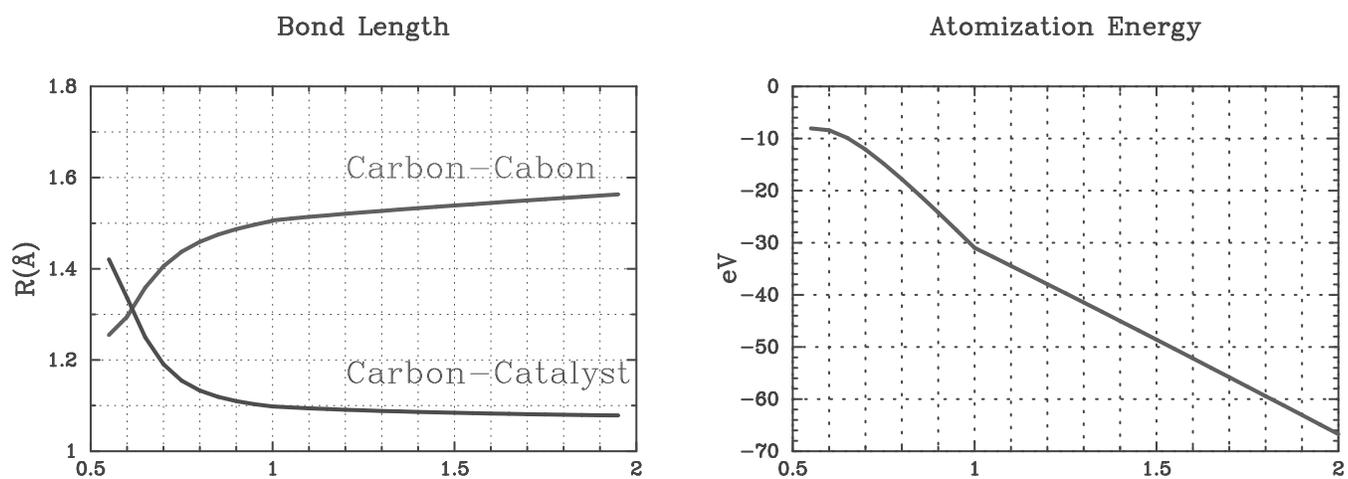
\item Figure 2: Parameter dependence of carbon-catalyst bonding
                in a C$_2$H$_6$-like molecule.
                The bond angles are  little dependent on $\lambda$.

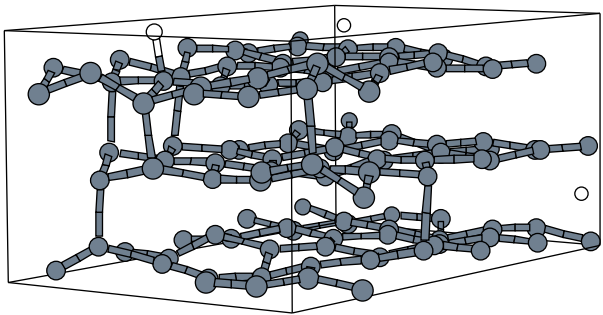
\item Figure 3: Buckling of graphite plane (gray sphere)
                 caused by catalyst (white sphere).

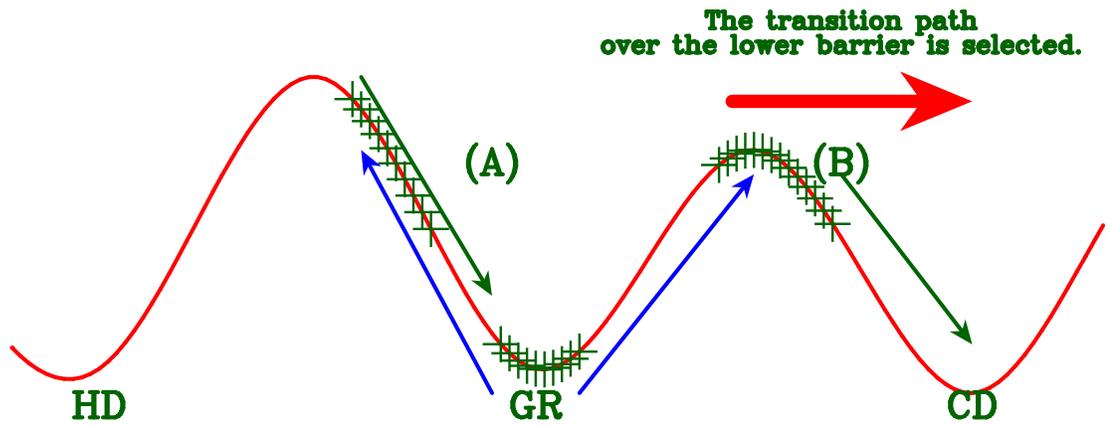
\item Figure 4: Activation barrier from graphite to diamond.

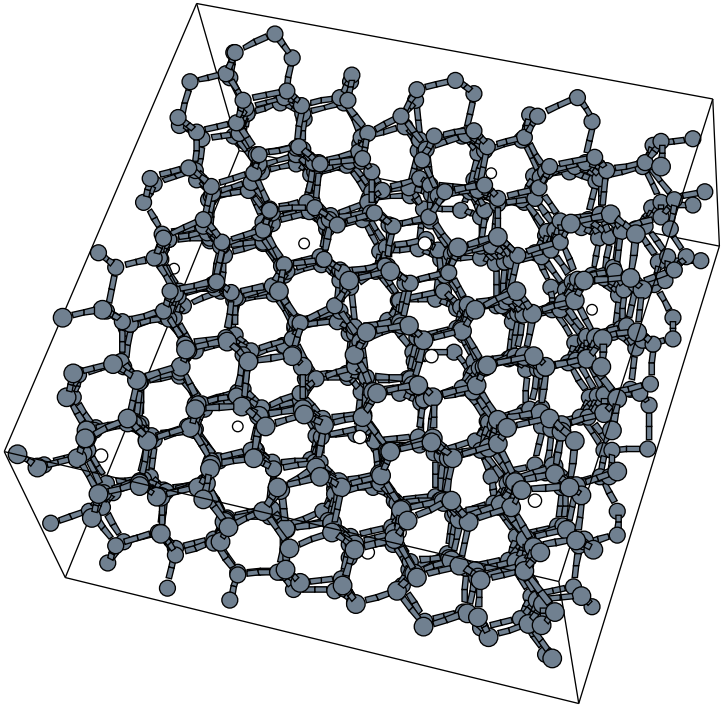
\item Figure 5: Hexagonal diamond structure 
                obtained by the simulation( 150GPa, 6000K, $\lambda=1.2$).

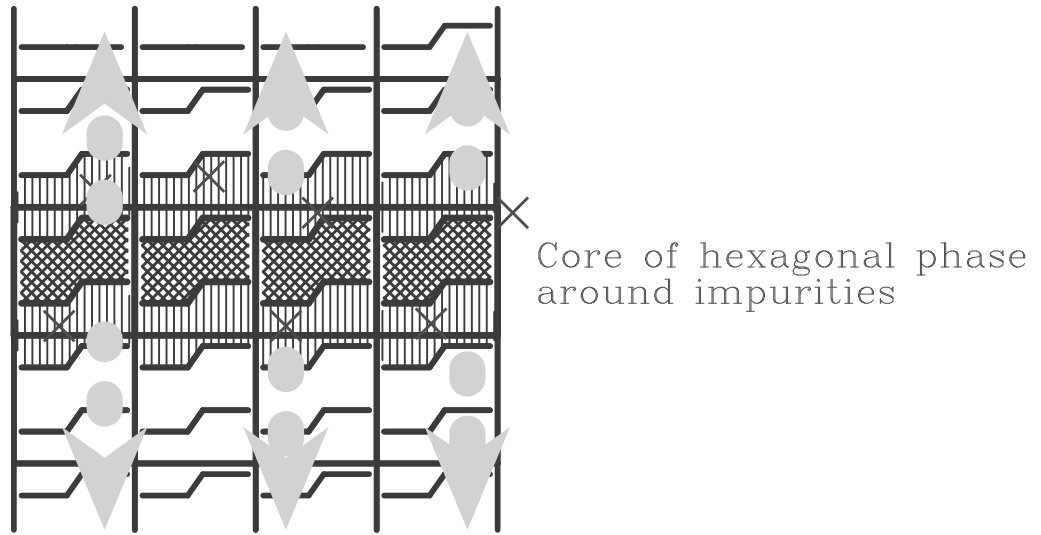
\item Figure 6: Formation of cubic and hexagonal diamond.
                (a) Growth of cubic diamond.(b) Growth of hexagonal diamond.

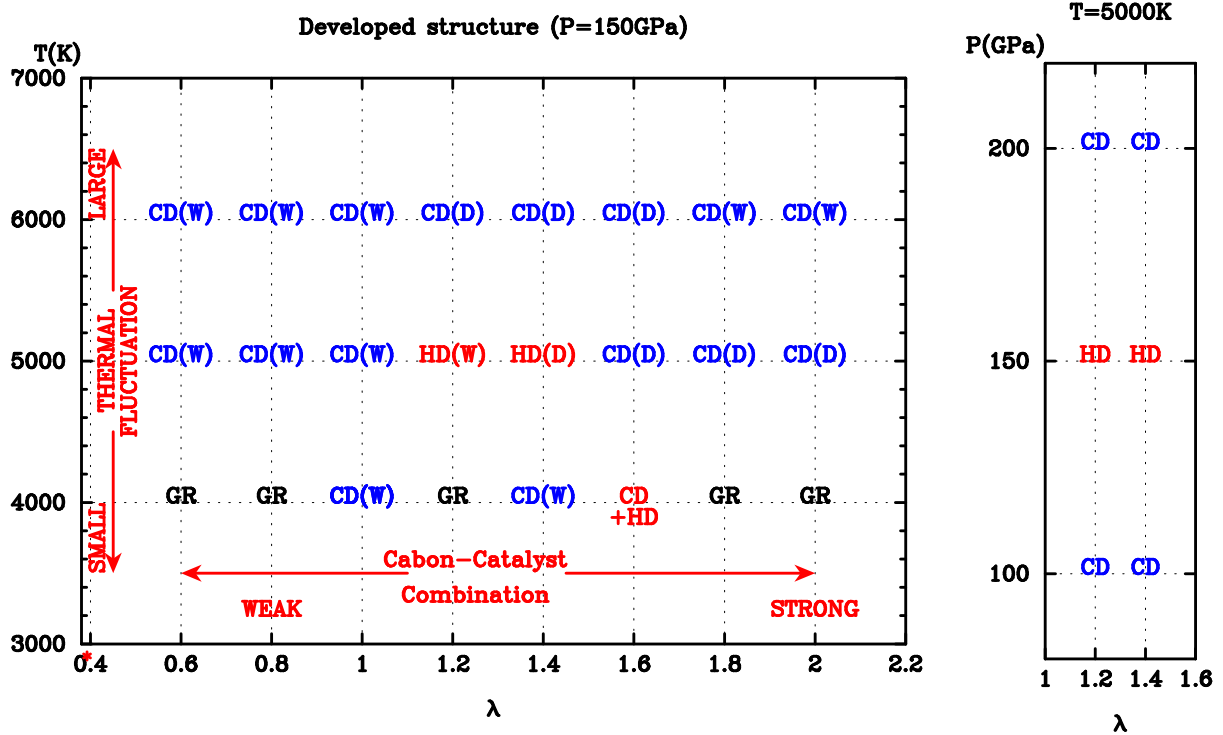
\item Figure 7: Synthesized phase after the compression 
                with various condition,  started from a certain
                initial geometry. (The initial geometry
                is given in figure 8.)
                The meaning of notations in the figure  is as follows.
                GR:Graphite,
                CD(W): Well-crystallized cubic diamond phase,
                CD(D): cubic phase with large defect,
                HD(W/D):Hexagonal phase, HD+CD:Coexistence of
                hexagonal and cubic phase.
                The transformed structures after MD simulations of 2ps 
                are shown there.

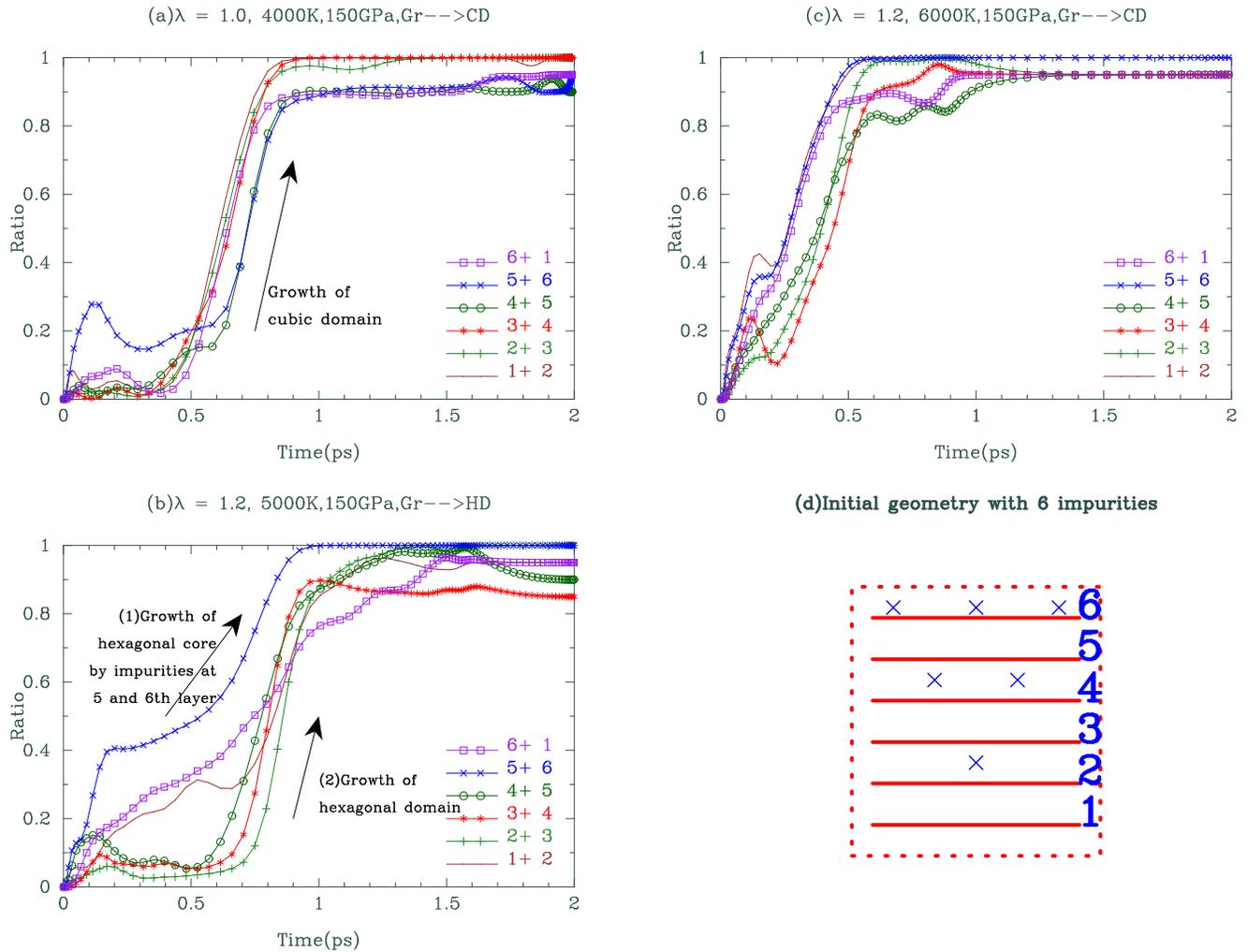
\item Figure 8: Formation of interlayer bonding
                with the presence of impurities.
                In each simulation,
                the pressure is increased from 0 to 150GPa
                in the first 0.15ps and kept constant after that.
                (a)-(c) :Plot with various conditions,
                each of which corresponds to 
                synthesized phases listed in figure 7.
                (d):Initial geometry with 6 impurities used
                   in the simulation of figure 7 and 8.
                   Broad lines and crosses show
                   the relative location of 6 graphite sheets
                   and 6 impurities in a cell.
                   One graphite sheet includes 40 C atoms.

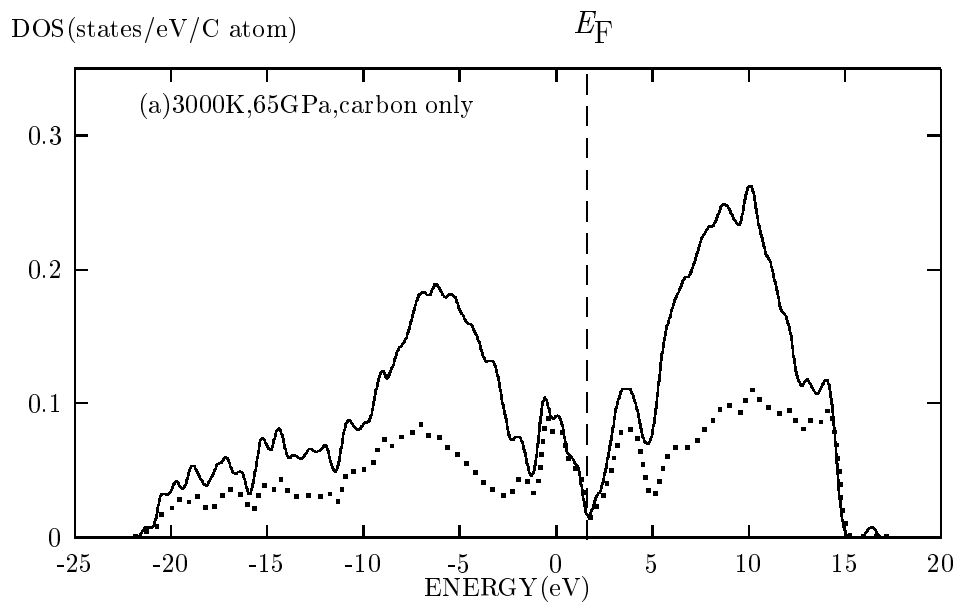
\item Figure 9: Density of states in amorphous structures
                synthesized by the compression of fullerene FCC crystal.
        (a): DOS in the quenched structure
             after the compression at 3000K, 65GPa.
        (b): after the compression at 5500K, 125GPa.
        (c): after the compression at 3500K, 55GPa,
             with the presence of 30 hydrogen atoms
             among 240 carbon atoms in the unit cell.
            In these figures, the real and dotted lines
            show the total DOS and the contribution from threefold 
            carbon, respectively.

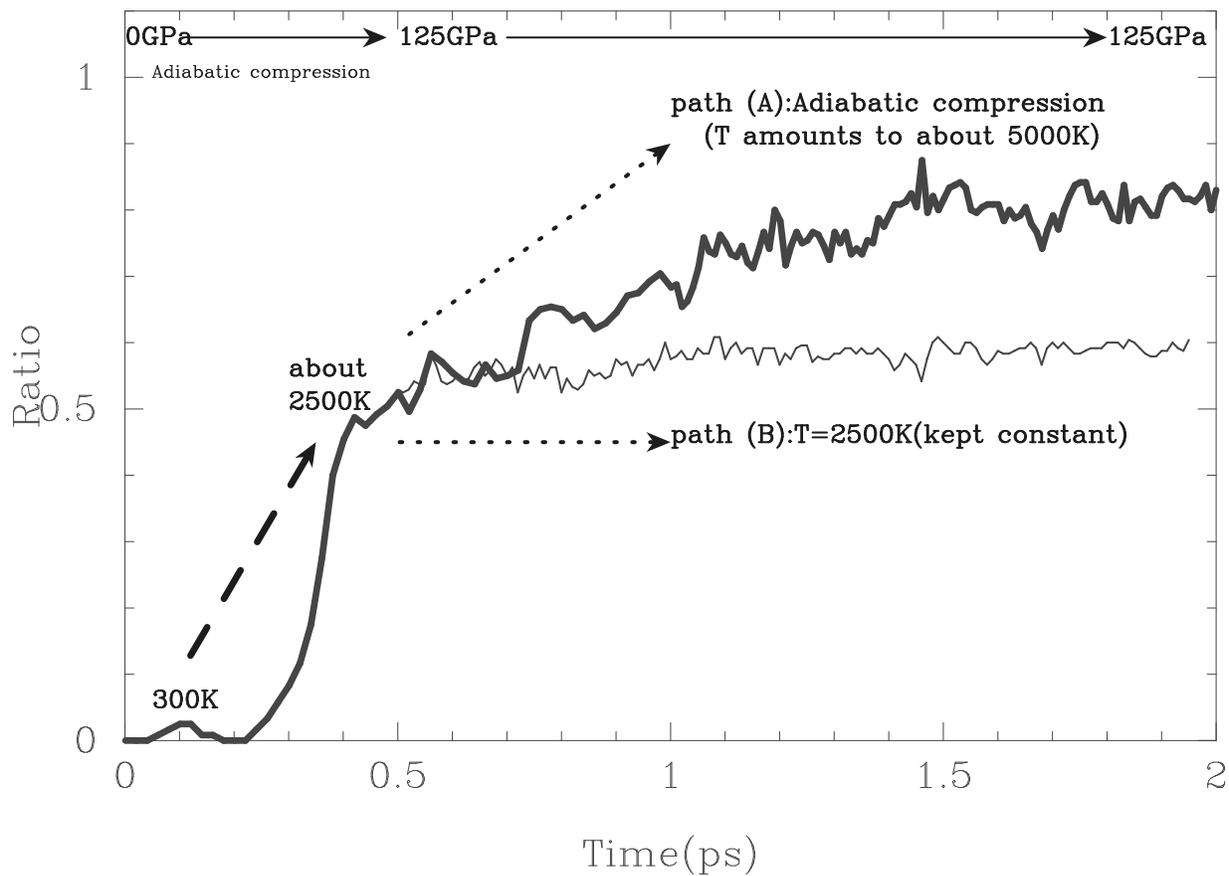
\item Figure 10: Formation of fourfold atoms
                 in the compression of fullerene fcc cell.
                 This figure shows the change in the ratio of 
                 fourfold carbon thorough the compression.
                 In path (A), the simulation is executed adiabatically
                 and the temperature increases to about 5000K.
                 In path (B), the simulation is executes adiabatically
                 at first, and afterwards, the temperature is kept
                 2500K.

\end{itemize}

\begin{thebibliography}{1}
\bibitem{BUNDY_KASPER}F.P.Bundy and J.S.Kasper, J.Chem.Phys. {\bf 46},
3437(1967).
\bibitem{CLARKE}R.Clarke and C.Uher, Adv.Phys. {\bf 33},469(1984).
\bibitem{1STP_CAL}S.Scandolo,M.Bernasconi, G.L.Chiarotti,P.Focher, and 
   E.Tosattti, Phys.Rev.Lett.{\bf 74} 4015(1995).
\bibitem{POLYGRA}S.Endo, N.Idani, R.Oshima, K.J.Takano and
M.Wakatsuki, Rhys. Rev. B , {\bf 49}, 22(1994).
\bibitem{YAGI}T.Yagi,W.Utsumi,M.Yamataka,T.Kikegawa, and O.Shimomura,
Phys.Rev. B {\bf 46}, 6031(1992)

\bibitem{N_DIA} H.Hirai and K.Kondo, Science {\bf 253}, 772(1991),
H.Hirai, K.Kondo, H.Sugiura, Appl. Phys. Lett. {\bf 61}, 27(1002).

\bibitem{G2D} Y.Tateyama, T.Ogitsu, K.Kusakabe, and S.Tsuneyuki,
              Phys.Rev.B{\bf 54},14994(1996).

\bibitem{Wentz} R.M.Wentzcovitch, J.L.Martins, and G.D.Price,
               Phys.Rev.Lett.{\bf 70}, 3947(1993);
               R.M.Wentzcovitch, W.W.Schulz, and P.B.Allen,
               {\it ibid.}{\bf 75},3389(1994).

\bibitem{PARAM_CH}M.D.Winn,M.Rassinger, and J.Hafner,
                  Phys.Rev.B{\bf 55},5364(1997).


\bibitem{HIRAI} H.Hirai, K.Kondo, N.Yoshizawa, and M.Shiraishi,
                Appl.Phys.Lett.{\bf 64} 1797(1994);
                H.Hirai and K.Kondo, Phys.Rev.B{\bf 51},15555(1995);

\bibitem{C60FCCBAND} S.Saito and A.Oshiyama, Rhys.Rev.Lett.{\bf 66}, 
                    2367(1991).

\bibitem{Utsumi}W.Utsumi and T.Yagi, Science {\bf 252}, 1542(1991).
\bibitem{LI_INT}S.Tsuneyuki,T.Ogitsu, Y.Tateyama, K.Kusakabe and
A.Kikuchi, \'Advances in High pressure Research in Condensed
Matter (Proceedings of the International Conference on
Condensed Matter under High Pressure(Bombay, India,11-15 November 1996)\'
(NISCOM,New Delhi, India 1997), pp.104-108.
\end{thebibliography}
\end{document}